# Renal Blood Flow Quantification During Standard Myocardial Perfusion Imaging with Rubidium-82 Positron Emission Tomography

Hamid Abdollahi[1,2], Robert Doot[3], Raul Porto,[3] Anthony Young,[3] Abdel K Tahari,[4] Raymond Townsend,[5] Jacob Dubroff[3], Arman Rahmim[1,2], Paco E. Bravo[3,6]

**Author affiliations:**

[1]Department of Radiology, University of British Columbia, Vancouver, BC Canada

[2]Department of Basic & Translational Research, BC Cancer Research Institute, Vancouver, BC Canada

[3]Division of Nuclear Medicine, Department of Radiology, Perelman School of Medicine, University of Pennsylvania, Philadelphia, PA, USA.

[4]Department of Clinical Imaging Tawam Hospital/Johns Hopkins International and College of Medicine and Health Sciences Al Ain, Abu Dhabi, United Arab Emirates.

[5]Division of Renal, Electrolyte and Hypertension, Department of Medicine, Perelman School of Medicine, University of Pennsylvania, Philadelphia, PA, USA.

[6]Division of Cardiovascular Medicine, Department of Medicine, Perelman School of Medicine, University of Pennsylvania, Philadelphia, PA, USA.

**Address for correspondence:**
Paco Bravo, MD
Hospital of the University of Pennsylvania
3400 Civic Center Blvd, 11-154 South Pavilion
Philadelphia, PA 19104
Tel: 1-215-220-9494
E-mail: paco.bravo@pennmedicine.upenn.edu

**ABSTRACT**

**Background:** Renal blood flow (RBF) is a fundamental marker of kidney health, yet noninvasive assessment is rarely incorporated into routine clinical imaging. We evaluated the feasibility and physiologic validity of quantifying renal transport of Rubidium-82 ($K_1$) during standard myocardial perfusion imaging (MPI) PET.

**Methods:** We studied 126 patients (age 60 ± 12 years; 48% male; 51% Black) undergoing clinically indicated rest/stress $^{82}$Rb MPI in whom at least one kidney was partially visualized within the axial field of view. Volumes of interest were drawn over the visible renal cortex. $K_1$ was estimated using a one-tissue compartment model with arterial input functions (AIF) derived from either the left ventricle (LV) or abdominal aorta.

**Results:** LV-derived AIF produced physiologic and internally consistent flow estimates, whereas aorta-derived AIF systematically overestimated $K_1$ and flow. LV-based measurements were therefore used for all analyses. $K_1$ demonstrated nonlinear flow dependence consistent with the Renkin–Crone extraction model, plateauing at higher perfusion states. Renal $K_1$ and flow declined progressively with worsening kidney function, from 1.24 ± 0.35 ml/min/g (eGFR ≥60) to 0.53 ± 0.22 ml/min/g (eGFR <15; P<0.0001). Only patients with preserved eGFR exhibited significant hyperemic augmentation. ROC analysis demonstrated excellent discrimination for reduced kidney function (AUC >0.90).

**Conclusion:** Opportunistic renal $K_1$ quantification during routine $^{82}$Rb PET is feasible, physiologically coherent, and strongly associated with kidney function.

## INTRODUCTION

Positron emission tomography (PET)/computed tomography (CT) is a robust modality for myocardial perfusion imaging (MPI). In addition, list-mode technology allows rapid, multi-frame dynamic acquisitions of the radiotracer activity in tissue and blood pool to generate time activity curves (TAC) and the arterial input function (AIF) respectively, which can then be applied to radiotracer- and tissue-specific kinetic models for absolute quantification of blood flow. The positron emitters, Oxygen-15 ($^{15}$O)-water, Nitrogen-13 ($^{13}$N)-ammonia, and Rubidium-82 ($^{82}$Rb) are particularly suitable for myocardial blood flow (MBF) quantification. While blood flow quantification in organs other than the heart (e.g. brain and kidney) is clinically feasible,(1, 2) simultaneous multi-organ blood flow quantification remains challenging due to the limited axial field of view (FOV) of most PET/CT systems. However, the advent of longer axial FOV PET cameras,(3) makes the possibility of simultaneous multi-organ blood flow quantification appealing, which may be helpful in order to gain insights into the mechanisms of disease of interrelated systems.

In the present study we aimed to investigate the feasibility of simultaneous MBF and renal blood flow (RBF) quantification in patients referred for clinically indicated pharmacologic stress testing with $^{82}$Rb PET/CT. Specifically, we sought to 1) measure transport (K1) of $^{82}$Rb into the kidneys (as surrogate marker of RBF) and 2) correlate renal K1 with clinical parameters. We hypothesize that this novel approach may be particularly suited to the investigation of individuals with suspected or known reno-vascular disorders.

## MATERIALS AND METHODS

### Patient Population

We retrospectively reviewed all images of patients who underwent myocardial perfusion pharmacologic stress test with $^{82}$Rb PET/CT for clinical indication at the Hospital of the University of Pennsylvania within a 6 month span (n=399) and included those patients in whom the right and/or left kidney were partial or fully included in the axial FOV of the heart (Figure 1). We also reviewed clinical history and relevant recent laboratory data. CKD-EPI Creatinine Equation was used to estimate glomerular filtration rate (GFR). The study was approved by the University of Pennsylvania Institutional Review Board with a waiver for informed consent.

**PET protocol**

Patients underwent rest-stress myocardial perfusion imaging with $^{82}$Rb on either a Biograph mCT (n=173) PET/CT (axial FOV 22.1 cm) or a Biograph Vision (n=225) PET/CT scanner (axial FOV 26.3 cm) (Siemens Healthineers, Malvern, PA). After an initial CT scout to delineate the FOV and localize the heart position, a single low-dose CT scan was acquired for attenuation correction (AC) of the subsequent PET emission scan. Subsequently, rest PET images were obtained with an approximately 6-minute list-mode PET acquisition imaging while 20 – 30mCi of $^{82}$Rb were injected intravenously as a bolus. Then, a standard intravenous infusion of regadenoson was given and a second dose of $^{82}$Rb (20 – 30mCi of $^{82}$Rb) was injected at peak hyperemia. Stress PET images were recorded in the same manner. List-mode data were resampled to static (90-second prescan delay), gated (8 bins per cardiac cycle), and dynamic image sets with a range from 21- to 28-frames for both rest and stress datasets.

**Image Segmentation**

For extraction of regional $^{82}$Rb activity, both static and dynamic PET datasets were uploaded simultaneously into commercial software (MIM v6.9a Software Inc., Cleveland, OH) followed by delineation of volumes of interest (VOIs). Manual contouring of the entire left ventricular (LV) wall and visible portion(s) of the right and/or left kidney(s) was performed on the axial plane of the static images, excluding the unreliable lowermost 2 slices of image data. Adequate VOI position was confirmed on the sagittal and coronal PET planes as well as the corresponding AC-CT. Blood activity concentration was measured in the left ventricular chamber as well as the descending aorta at the level of L1 with approximately 1.5 mL spherical VOIs (Suppl. Figure 1). All VOIs were automatically applied to the dynamic datasets to allow extraction of tissue specific TACs. This process was repeated on the stress PET/CT datasets.

**Image Processing**

Kinetic modeling of $^{82}$Rb flow was performed in MATLAB. For renal flow quantification, extracted kidney TACs were fitted using a 1-tissue compartment model (also called a two-compartment model with a vascular and a cellular compartment).(2) The AIF was derived from the LV blood pool and abdominal aorta and fitted using a tri-exponential function. Compartmental model fitting was performed with blood delay, using with a fixed fractional blood volume based on Yamashita's work(4). For MBF quantification, TACs-derived from the LV were fitted using a standard 1-tissue-compartment model.(5) (6) Coronary flow reserve (CFR) was defined as the ratio of hyperemic/stress to resting MBF.

**Microparameter Estimation ($v_B$, $K_1$, $k_2$)**

Renal tracer kinetics were modeled using a one-tissue compartment model incorporating a vascular fractional volume term.(2) The measured renal activity concentration $C_T(t)$ was expressed as:

$$C_T(t) = (1 - v_B)\, C_{tis}(t) + v_B\, C_p(t) \qquad (1)$$

where $v_B$ represents fractional blood volume within the renal cortical VOI, $C_p(t)$ is the arterial input function (AIF), and $C_{tis}(t)$ represents tracer concentration within the exchangeable renal tissue compartment. Tissue kinetics were governed by:

$$\frac{dC_{tis}(t)}{dt} = K_1 C_p(t) - k_2 C_{tis}(t) \qquad (2)$$

where $K_1$ (mL/min/g) reflects tracer transport from blood to tissue (flow × extraction fraction), and $k_2$ (min$^{-1}$) represents tracer efflux from tissue back to blood.

Rather than numerically solving the differential equation at each iteration, a closed-form convolution solution was implemented in MATLAB:

$$C_{tis}(t) = K_1 e^{-k_2 t} \int_0^t C_p(s) e^{k_2 s} ds \qquad (3)$$

Model fitting was performed using constrained nonlinear optimization (interior-point algorithm), minimizing the sum of squared residuals between measured and modeled tissue TACs:

$$\text{SSE} = \sum_{i=1}^{N} \left[ C_T(t_i) - \hat{C}_T(t_i) \right]^2 \qquad (4)$$

Physiologic bounds were imposed ($0 \leq v_B \leq 1$, $K_1 \geq 0$, $k_2 \geq 0$). Residual analysis and percent error metrics were used to confirm goodness-of-fit. The resulting fitted parameters $v_B$, $K_1$, and $k_2$ formed the basis for subsequent renal blood flow quantification.

**Renal Blood Flow (F) Quantification from $K_1$**

Because $K_1$ represents the product of renal blood flow (F) and tracer extraction fraction (E), absolute flow was derived according to:(7)

$$K_1 = F \cdot E \qquad (5)$$

If extraction were constant, flow could be directly estimated as:

$$F = \frac{K_1}{E} \qquad (6)$$

However, consistent with prior experimental and human data demonstrating flow-dependent extraction of $^{82}$Rb, we applied a Renkin–Crone formulation:

$$K_1 = F\left(1 - a e^{-b/F}\right) \qquad (7)$$

where $a$ and $b$ represent permeability–surface area parameters of renal microvasculature. For each fitted $K_1$, flow $F$ was obtained by numerically solving the nonlinear equation above. Final flow values were expressed in mL/min/g assuming a renal tissue density of 1.04 g/mL.

This approach preserves the known saturable extraction behavior of $^{82}$Rb, ensuring physiologic coherence across low and high perfusion states and enabling direct comparison with established renal perfusion literature.

**Standard Myocardial Perfusion Imaging**

Static and gated PET MPI scans were analyzed using commercially available software (Corridor4DM; Ann Arbor, Michigan). Abnormal MPI was defined as a summed stress score ≥ 2. Left ventricular ejection fraction (LVEF) was calculated from gated images (8-frame per cardiac cycle).

**Statistical Analysis**

Paired t test was used to evaluate statistical differences between 2 continuous measurements or variables of the same individuals. An independent-measures t-test was used to assess continuous variables differences between 2 groups. One-way ANOVA was performed to compare mean values of >2 groups. Continuous variables are presented as mean ± SD. Categorical variables were compared between groups using $\chi^2$ tests and are presented as percentages. Sensitivity, specificity, positive predictive value (PPV), negative predictive value (NPV) and the area under the receiver operating characteristic (ROC) curve were computed for renal K1 using GFR < 30 mL/min/1.73 m$^2$ as the reference standard. Simple correlations were assessed using Pearson's coefficient correlation. A P value <0.05 was considered statistically significant. All statistical analyses were performed using Stata version 13 (StataCorp LLC, College Station, Texas).

**RESULTS**

We identified a total of 145 patients (36%) that were initially eligible, however, after excluding patients with incomplete datasets, technical reasons, and/or renal volumes of less than 20 mL (to minimize the contribution of partial volume effect), the final study cohort consisted of 126 patients (32%), including 86, 25 and 15 patients in whom either both kidneys, the left kidney only, or the right kidney only were included in the FOV, respectively.

Baseline characteristics of the final cohort are depicted in Table 1. Approximately 50% of individuals were Black, and most patients had associated comorbidities, including end-stage renal disease (ESRD) in 30% of the study population.

**Arterial input function contribution: LV versus abdominal Aorta**

In direct comparison, both K1 and flow estimates were systematically higher when derived using the aorta-based AIF than when using the LV-based method, at rest and during vasodilator stress (**Table 2 and Figure 2**). LV-derived measurements produced flow values that remained within physiologically expected renal perfusion ranges, particularly at baseline. In contrast, aorta-derived values frequently exceeded the known upper limits of renal blood flow, indicating systematic overestimation of tracer delivery when the aorta is used as the AIF (**Table 2 and Figure 2**).

When comparing flow between kidneys, renal K1 and flow estimates were closely comparable between the right and left kidneys at baseline and during vasodilator stress for both AIF (**Table 2 and Figure 3**). The mean right–left renal K1 differences consistently clustered around zero across all conditions, indicating no systematic bias in either direction (**Figure 3**). Variability was modestly greater during vasodilator stress than at rest, reflecting expected physiologic heterogeneity at higher flow states. Although both AIF methods demonstrated reasonably narrow limits of agreement, LV-derived K1 values showed tighter dispersion and narrower limits of agreement compared with their aorta-derived counterparts. This suggests that the LV-derived AIF provides more internally consistent bilateral renal K1 estimates. Taken together, these findings support the LV-derived AIF as the more robust and physiologically reliable method for quantifying renal perfusion with Rb-82 PET.

**Renal flow metrics: K1 and Flow**

Using the LV-derived AIF, renal K1 demonstrated a strong nonlinear relationship with flow at both rest and vasodilator stress, following the expected saturable extraction behavior (**Figure 2A**). Stress conditions shifted the curve upward, reflecting higher absolute flow and K1 values. When K1 was expressed as extraction fraction (K1/F), values were highest at low flows and declined progressively with increasing perfusion. This inverse relationship was nearly identical between rest and stress, indicating that the flow-dependent extraction behavior of Rb-82 is preserved across physiologic states. Together, these findings confirm that LV-derived Rb-82 kinetics yield physiologically coherent K1–flow relationships and extraction fraction patterns consistent with classic tracer extraction models.

**Renal flow and Kidney Function**

Renal K1 and flow decreased in proportion to the reduction in eGFR, and only patients with preserved eGFR demonstrated a significant flow increase in response to vasodilator stress (**Figure 4**).

In a direct comparison of the two metrics, K1 exhibited a stronger linear correlation with serum creatinine than flow, resulting in higher Pearson r values (**Figure 5**). However, Spearman rho values for K1 and flow were nearly identical consistent with a similarly strong monotonic increase across the full range of eGFR.

In ROC analysis, LV-derived K1 and flow demonstrated nearly identical diagnostic performance for identifying reduced kidney function (**Table 3**). All AUC values exceeded 0.90, indicating excellent ability to detect different degrees of impaired eGFR. Diagnostic performance improved progressively when evaluating more severe renal dysfunction, with the highest AUCs observed for identifying patients with GFR less than 30. This pattern was consistent for both K1 and flow. Paired ROC comparisons showed no significant differences between the two metrics at any threshold. These findings indicate that LV-derived K1 and flow provide equivalent discrimination across clinically relevant stages of reduced eGFR.

**Differences between scanners**

Only 6 patients had complete visualization of the kidneys (all in the Vision scanner), whereas the remaining 120 patients had partial visualization of either kidney (**Figure 1**). The visualized volumes of the right (84 ± 46 vs. 52 ± 32 mL; P=0.0007) and left kidneys (79 ± 41 vs. 53 ± 25 mL; P=0.0014) were significantly greater in the Vision compared to the smaller axial FOV mCT PET/CT system. No significant differences were seen on the renal K1 estimates obtained on the mCT compared to the Vision scanner at rest (P=0.15) or during vasodilator stress (P=0.24).

**DISCUSSION**

The present study demonstrates that renal perfusion metrics can be reliably computed from routine Rb-82 myocardial perfusion PET studies and that these measurements behave in a physiologically manner across the full spectrum of kidney function. By leveraging large scale clinical cardiac PET data, this work bridges the gap between controlled experimental studies of Rb-82 renal kinetics and real-world clinical practice. In doing so, it provides new methodological clarity regarding AIF selection, extends extraction limited kinetic principles into a diverse clinical population, and establishes strong associations between metrics of renal flow and kidney functional status. The findings build upon and refine several years of Rb-82 PET research and offer a practical framework for opportunistic cardiorenal imaging.

**Arterial Input Function Selection: LV Versus Aorta**

Prior human studies of renal Rb-82 PET have generally accepted the abdominal aorta as an adequate arterial input function.(2, 7-9) Langaa and colleagues reported similar K1 estimates from LV and aortic AIF in 18 healthy volunteers in a single FOV Rb-82 PET/CT acquisition.(8) Similarly, Van de Burgt found no significant differences in RBF calculation using the aorta- and LV-based AIFs among 17 patients undergoing myocardial perfusion PET.(7) Moreover, Bibeau-Delisle et al, reported aorta-derived RBF values that ranged from 1.32 to 14.95 and 0.78 to 8.29 ml/min/g in patients with normal and abnormal renal function respectively.(9)

Our findings challenge the generalizability of this assumption. In a real-world cardiac population in whom renal coverage is often partial, aorta-derived AIF consistently overestimated tracer delivery and produced K1 and flow values that frequently exceeded known physiologic limits of human renal perfusion.(10-14) Respiratory motion, partial volume spill-in and bolus dispersion within the abdominal aorta are potential factors that contributed to these inflated measurements. In contrast, LV-based input functions produced stable and physiological curves across the entire cohort and generated bilateral K1 estimates with tighter agreement. These observations suggest that the LV cavity is a more reliable and physiologic input function for opportunistic renal flow quantification during routine cardiac PET imaging.

**Flow Estimation from K1**

Foundational work by Tamaki and Mullani established that Rb-82 exhibits a high first-pass extraction fraction and that uptake in renal cortex is governed by extraction limited kinetics.(15, 16) Later human studies by Gregg and colleagues demonstrated that renal K1 follows a nonlinear relationship with flow consistent with the Crone Renkin extraction model.(17)

Our results reaffirm these principles in a clinical population undergoing routine cardiac PET. Using an LV-based input function and a validated extraction curve, we observed the expected saturable K1 to flow relationship, with high extraction fractions at low perfusion and declining extraction at higher flows. Rest and stress relationships overlapped, confirming preserved saturable behavior across physiological states. These parallels indicate that even with partial kidney visualization, Rb-82 transport parameters derived from cardiac PET datasets retain the physiological behavior documented in controlled experimental settings.

**Association of Renal Flow Metrics with Kidney Function**

Several studies have reported associations between Rb-82 renal kinetics and kidney function. Bibeau Delisle observed that renal flow and resistance differ between normal and abnormal eGFR groups, while Langaa demonstrated that Rb-82 clearance approximates MAG3 clearance.

Our findings extend these observations by evaluating a large population spanning the entire spectrum of kidney function, including many individuals with advanced chronic kidney disease. Both K1 and model derived flow showed progressive decline across ordered eGFR categories and strong monotonic relationships with serum creatinine. Hyperemic responses were restricted to participants with preserved eGFR, reinforcing prior observations of early impairment in renal vasodilatory reserve. Importantly, K1 and flow demonstrated excellent diagnostic performance for identifying clinically relevant eGFR thresholds, with area under the curve values consistently greater than 0.90. The comparable diagnostic performance of K1 and flow suggests that K1 alone may serve as a practical surrogate for renal perfusion, simplifying implementation in clinical settings.

**Overall Significance**

By demonstrating physiologic validity, methodological robustness and strong clinical associations, this study positions renal Rb-82 quantification as a feasible and informative extension of routine cardiac PET imaging. In contrast to earlier work reliant on full renal coverage or controlled acquisition environments, our findings show that reliable renal perfusion metrics can be obtained opportunistically during standard MPI, if using LV derived input functions. These insights expand the applicability of renal Rb-82 PET and support its potential role in improving functional assessment in populations where cardiorenal interactions are central to disease progression.

## CONCLUSION

Quantification of renal K1 from standard Rb-82 cardiac PET is feasible, physiologically meaningful, and strongly correlated with kidney function. LV-derived AIFs provide the most reliable input for renal kinetic modeling, while aorta-derived AIFs systematically overestimate flow. Renal K1 follows classic extraction-limited tracer behavior, correlates with serum creatinine and eGFR, and demonstrates excellent discriminatory capacity for reduced kidney function. These findings establish the foundation for incorporating opportunistic renal perfusion assessment into routine cardiac PET workflows and support future investigation of its clinical role in renovascular and parenchymal kidney disease.

## DISCLOSURES

The authors report no relevant conflicts of interest.

## FUNDING SOURCE

Partial support was possible from a grant from the University of Pennsylvania Research Foundation.

**TABLE 1.** Baseline characteristics

| Characteristic | $n = 126$* |
|---|---|
| Age, years | 60 ± 12 |
| Males, n | 61 (48) |
| Whites, n | 50 (40) |
| Blacks, n | 64 (51) |
| Body mass index, kg/m$^2$ | 34.1 ± 8.9 |
| Creatinine, mg/d | 1.11 [0.87 – 4.33] |
| GFR ≥ 60 mL/min/1.73 m$^2$, n | 70 (56) |
| GFR 45 – 59 mL/min/1.73 m$^2$, n | 9 (7) |
| GFR 30 – 44 mL/min/1.73 m$^2$, n | 4 (3.2) |
| GFR 15 – 29 mL/min/1.73 m$^2$, n | 5 (4) |
| GFR < 15 mL/min/1.73 m$^2$, n | 38 (30) |
| Dialysis, n | 27 (21) |
| Hypertension, n | 105 (83) |
| Diabetes, n | 58 (46) |
| Coronary artery disease, n | 31 (25) |
| Heart failure, n | 10 (8) |
| Abnormal myocardial perfusion imaging, n | 29 (23) |
| Left ventricular ejection fraction, % | 61 ± 13 |
| Left ventricular ejection fraction < 40% | 11 (9) |
| Stress myocardial blood flow, mL/min/g | 1.82 ± 1.04 |
| Rest myocardial blood flow, mL/min/g | 0.98 ± 0.45 |
| Coronary flow reserve, unitless | 1.96 ± 1.03 |
| Heart rate, bpm | 72 ± 12 |
| Systolic blood pressure, mmHg | 139 ± 21 |
| Diastolic blood pressure, mmHg | 78 ± 11 |

* Data are expressed as mean ± standard deviation or median [Interquartile range] or as number (percentage).

GFR = Glomerular filtration rate.

TABLE 2. Renal volume and flow parameters when both kidneys were visible (volume of each kidney ≥ 20mL) in the field of view (± standard deviation)

| Variable | Left kidney | Range | Right kidney | Range | P value |
|---|---|---|---|---|---|
| **LV blood as AIF** | | | | | |
| Rest K1 (ml/min/g) | 1.01 ± 0.45 | 0.23 – 1.97 | 1.00 ± 0.46 | 0.23 – 1.95 | 0.52 |
| Rest flow (ml/min/g) | 1.73 ± 1.55 | 0.23 – 6.85 | 1.69 ± 1.49 | 0.23 – 6.64 | 0.65 |
| Stress K1 (ml/min/g) | 1.09 ± 0.55 | 0.20 – 2.44 | 1.09 ± 0.55 | 0.15 – 2.40 | 0.95 |
| Stress flow (ml/min/g) | 2.31 ± 2.57 | 0.20 – 12.9 | 2.29 ± 2.48 | 0.14 – 12.3 | 0.89 |
| **Aorta as AIF** | | | | | |
| Rest K1 (ml/min/g) | 1.44 ± 0.74 | 0.28 – 3.72 | 1.40 ± 0.68 | 0.35 – 3.65 | 0.24 |
| Rest flow (ml/min/g) | 4.73 ± 6.00 | 0.27 – 32.4 | 4.22 ± 5.20 | 0.34 – 31.2 | 0.09 |
| Stress K1 (ml/min/g) | 1.48 ± 0.84 | 0.22 – 4.30 | 1.48 ± 0.81 | 0.20 – 4.19 | 0.94 |
| Stress flow (ml/min/g) | 5.51 ± 7.87 | 0.21 – 41.5 | 5.38 ± 7.37 | 0.19 – 39.8 | 0.68 |

**Table 3.** Diagnostic performance of LV-based K1 and flow to identify reduced eGFR

| Outcome | Predictor | AUC | P value |
|---|---|---|---|
| GFR < 30 mL/min/1.73m$^2$ | K1 | 0.94 | 0.71 |
| | Flow | 0.95 | |
| GFR < 45 mL/min/1.73m$^2$ | K1 | 0.95 | 0.64 |
| | Flow | 0.95 | |
| GFR < 60 mL/min/1.73m$^2$ | K1 | 0.91 | 0.58 |
| | Flow | 0.92 | |

**Figure 1.** Representative partial and full visualization of kidneys in the axial field of view of patients undergoing myocardial perfusion Rb$^{82}$ PET imaging.

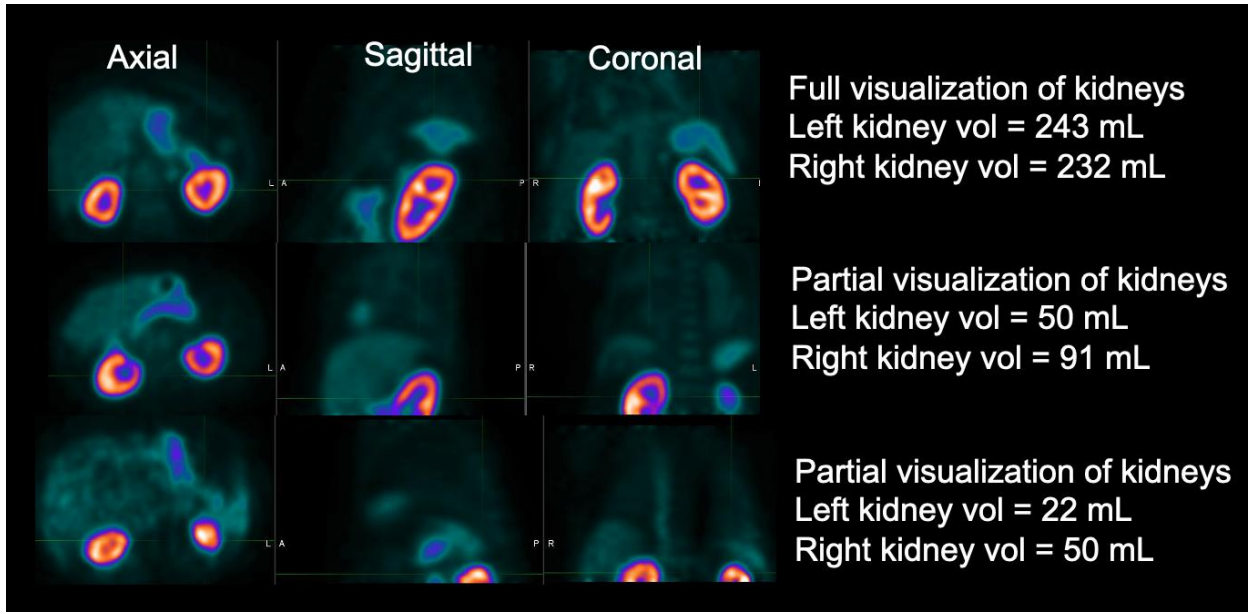

**Figure 2.** Comparison of rest and vasodilator stress $^{82}$Rb renal K1–Flow curves using LV-derived (A) and aorta-derived (B) arterial input functions.

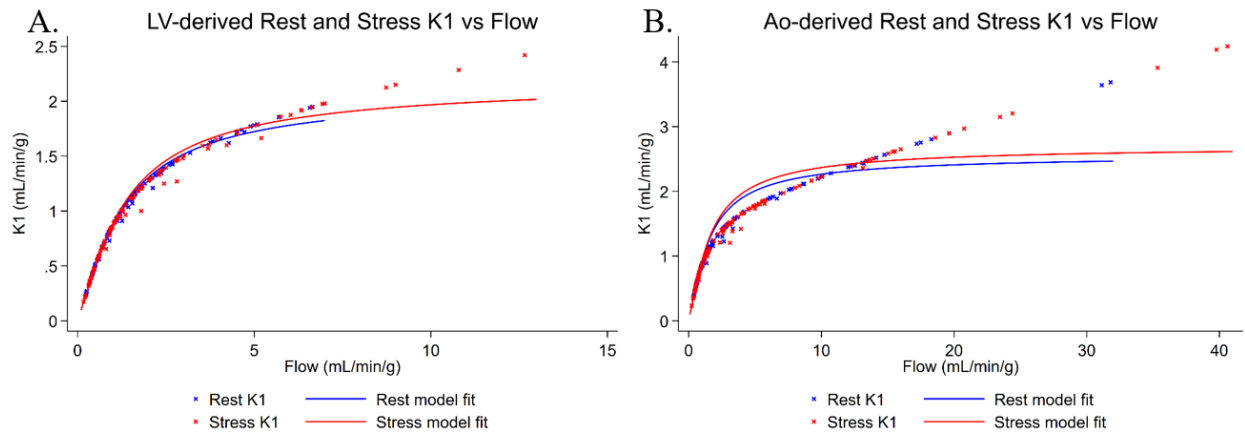

**Figure 3.** Agreement between right and left $^{82}$Rb renal K1 measurements using LV-derived (A) and aorta-derived (B) arterial input functions at rest and during vasodilator stress. Bland–Altman plots show the difference in K1 between right and left kidneys (y-axis) versus their mean K1 (x-axis). Filled black circles represent rest measurements and open black circles represent stress measurements. For each condition, the short-dashed black horizontal line denotes the mean bias at rest and the long-dashed black line denotes the mean bias during stress. Blue dashed lines indicate the rest limits of agreement (mean ± 1.96 SD) and blue long-dashed lines indicate the stress limits of agreement.

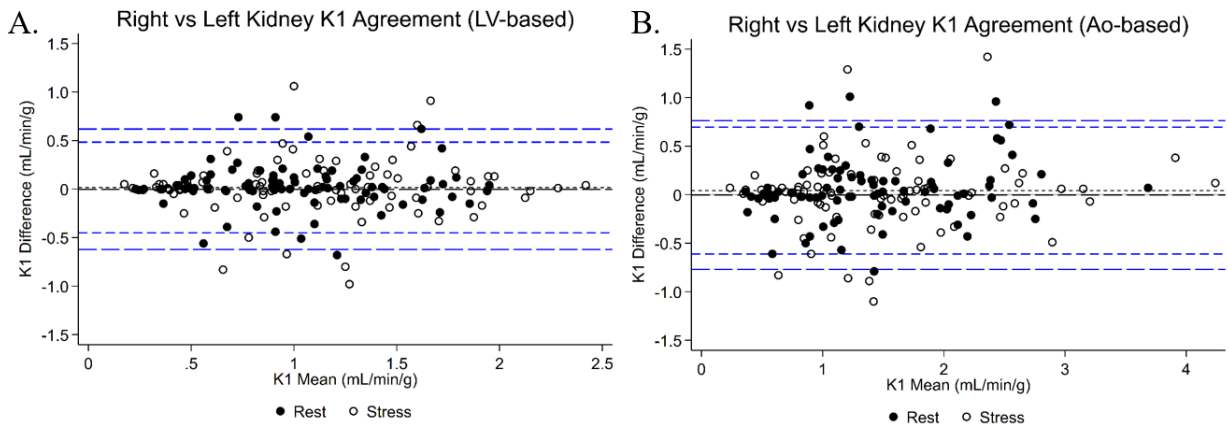

**Figure 4.** Proportional reduction of baseline and stress $^{82}$Rb renal K1 (A) and flow estimates (B) with worsening estimated glomerular filtration rate (eGFR). Renal K1/flow increased significantly in response to vasodilator stress only in the group with preserved eGFR.

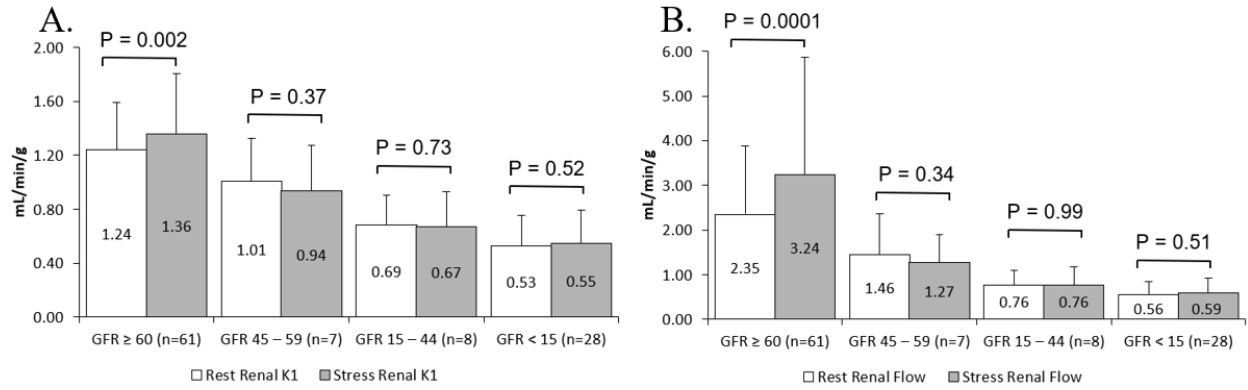

**Figure 5.** Association of $^{82}$Rb renal K1(A) and flow (B) with kidney function

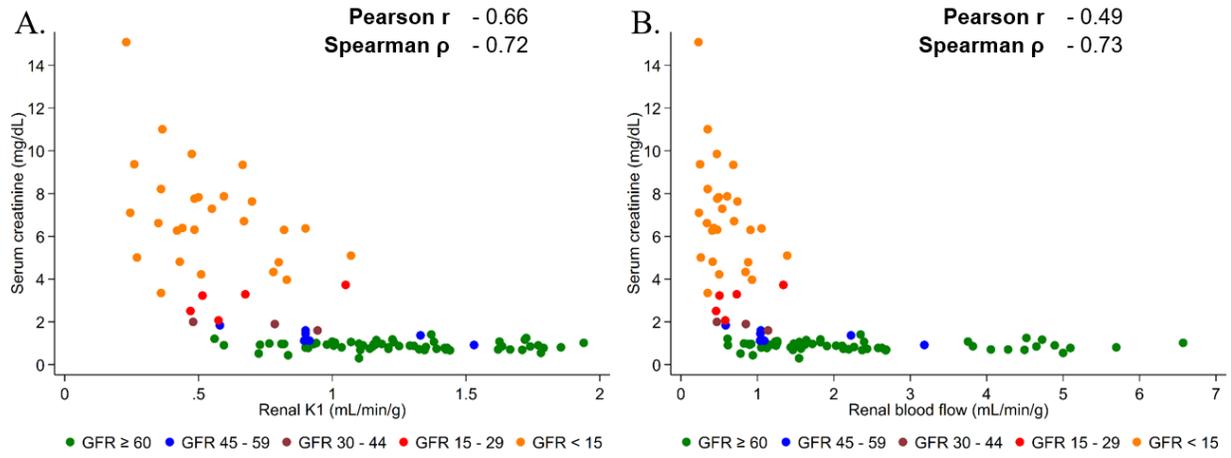